%% file: arXiv_FVscaling.tex
\documentclass[aps,pra,floatfix,twocolumn,superscriptaddress]{revtex4-2}

\usepackage{graphicx}
\usepackage{braket}
\usepackage{mathtools}
\usepackage{booktabs}

\input{preamble.tex}

\begin{document}
\preprint{}

\title{Universal Family--Vicsek scaling in quantum gases far from equilibrium}

\input{authors_FVscaling}


\begin{abstract}
Fluctuations in the growing surfaces of classical systems can exhibit universal scaling behavior, known as Family–Vicsek (FV) scaling. 
Although this phenomenon was originally discovered in classical stochastic models, recent theoretical studies have demonstrated the presence of FV scaling in quantum many-body systems as well.
Here, we observe the universal FV scaling in a one-dimensional Bose gas in an optical lattice.
By monitoring the fluctuations of particle number in half of the system, which corresponds to the surface roughness, we extract all scaling exponents and demonstrate that the entire relaxation—from the growth of quantum fluctuations to their saturation—is captured by a single universal scaling function. 
Our results demonstrate that universal scaling laws of classical surface growth extend to quantum many-body systems, establishing a unified framework for nonequilibrium universality across classical and quantum systems.
\end{abstract}
\maketitle


\section*{Introduction}
\begin{figure*}
\centering
\includegraphics[width=0.9\linewidth]{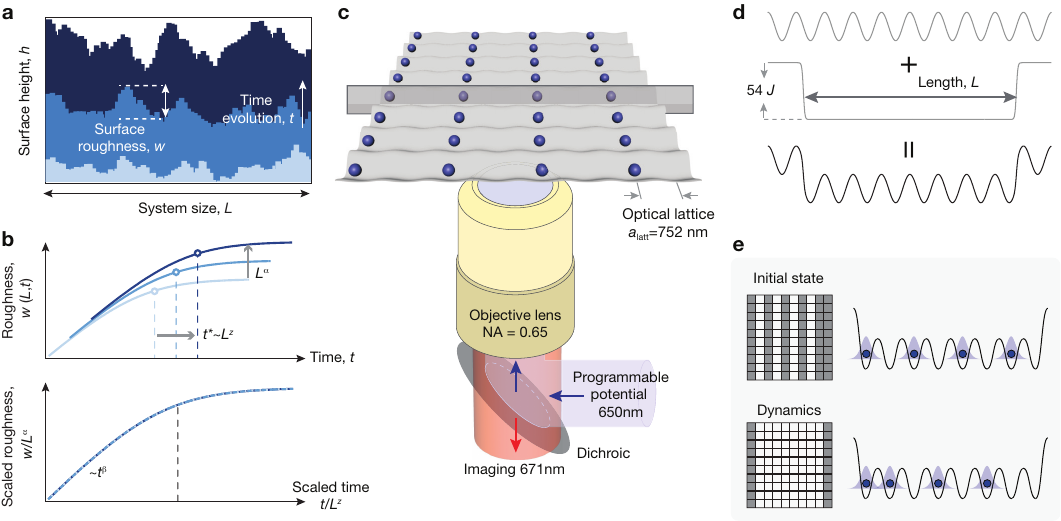}
\caption{\textbf{Family-Vicsek scaling and experimental scheme.}
\textbf{a,} Schematic illustration of surface growth dynamics in a one-dimensional system with a length $L$. 
As time $t$ evolves, the standard deviation of the surface height, called surface roughness $w(L,t)$, increases.
\textbf{b,} Time evolution of the surface roughness $w(L,t)$ at different system size. 
When we normalize the roughness and time by $L^{\alpha}$ and $L^z$, all curves follow a single universal function.
\textbf{c,} Quantum gas microscope setup for studying Family-Vicsek scaling in optical lattices. 
Using a high-resolution objective lens, we are able to image and manipulate atomic distribution with single-site resolution. 
The programmable potential is created by using a digital micro-mirror device (DMD) and is combined with the imaging beam path via a dichroic. 
\textbf{d,} Cross-section of the potential in the gray shaded region in (c). 
The repulsive addressing beam can create a barrier wall at the two ends with its maximum height~$54J$.
The experiments are carried out in the central region of the length $L$. 
\textbf{e,} Patterns in the addressing beam, where each square corresponds to an individual lattice site and shaded regions indicate sites where the light is turned on.
The staggered pattern is used to prepare the initial charge density wave state (top), while light is applied only to the outer edges during the subsequent dynamics (bottom).
}\label{ExpProtocol}
\end{figure*}

Universal scaling laws have provided a powerful framework for understanding and predicting emergent phenomena in many-body systems near equilibrium~\cite{Kadanoff1990,Cardy1996}. 
Such scaling behavior also arises in far-from-equilibrium systems, and the Family--Vicsek (FV) scaling~\cite{Vicsek1984,Family1985} stands out as one of the most fundamental laws in the stochastically growing surfaces. 
FV scaling characterizes how the roughness (fluctuations) of a growing surface evolves over time and scales with system size, and shows that the dynamics of surface roughness are governed by a small set of universal exponents, $\alpha,\beta$, and $z$ (Fig.~\ref{ExpProtocol}a and b). 
These exponents classify the nonequilibrium universality class of surface growth, where the Edwards–Wilkinson (EW)~\cite{Edwards1982} and Kardar--Parisi--Zhang (KPZ) models~\cite{Kardar1986,Amar1992} are the canonical examples. 
Originally developed in the context of a stochastic surface growth model, FV scaling has been observed in a wide range of systems from biological systems~\cite{Wakita1997,Huergo2010}, a fluid interface in a porous medium~\cite{Planet2007}, slow combustion of paper~\cite{Myllys2001},  turbulent liquid crystal~\cite{Takeuchi2010}, and sedimentation of particles in a fluid environment~\cite{Sardari2024}, deepening our understanding of the roughness dynamics~\cite{Barabasi2009,Takeuchi2018}.

While the dynamics of quantum many-body systems are fundamentally different from those of classical systems, the universal scaling has been widely observed in quantum systems far from equilibrium~\cite{Navon2014,Erne2018,Prufer2018,Glidden2021,Huh2024,Gazo2025}.
In particular, recent experiments on the dynamics of the integrable one-dimensional quantum spin chains have shown the KPZ-type superdiffusive dynamics even without external noise~\cite{Ljubotina2017,Ljubotina2019,Wei2022,Scheie2022,Rosenberg2024}.
While these works focused on dynamical scaling at the transient times alone, theoretical studies have further demonstrated the FV scaling in various closed quantum many-body systems by introducing a quantum-mechanical operator analog of the surface height~\cite{Fujimoto2020,Fujimoto2021,Fujimoto2022,Bhakuni2024,Aditya2024,Moca2025}, whose roughness growth arises from intrinsic quantum fluctuations.
That is, the time evolution of the roughness is collapsed into a single universal curve for all times by an appropriate spatiotemporal scaling characterized by universal exponents.
These findings suggest that the FV scaling offers a universal framework for describing nonequilibrium dynamics of both classical and quantum systems, yet experimental validation of the FV scaling in quantum many-body systems and the classification of their universality class remain to be established.

Here we report the observation of the FV scaling in an atomic one-dimensional Bose-Hubbard quantum simulator.
Using a high-resolution imaging system, we have succeeded in realizing a programmable site-addressing system, which can prepare a charge density wave (CDW) as an initial state and provides a systematic control of the chain length in one-dimensional lattices. 
In the first experiment, we investigate the dynamics in a strongly interacting regime with different system sizes.
A quantum surface height can be reconstructed from the site-occupation number and can quantify the surface roughness from its standard deviation. 
From the finite-size scaling experiments, we observe collapse of the surface roughness under spatiotemporal scaling and extract all dynamical exponents, $(\alpha,\beta,z)_{\rm exp}=[0.49(1),0.41(1),1.00(4)]$, where the exponents align with those expected for the ballistic universality class~\cite{Fujimoto2020}.
 
We further demonstrate that the nature of scaling behavior can be changed upon introducing a temporal disorder potential.
It is time-dependent disturbances in the form of on-site repulsive potentials that are randomly switched on in time.
Even in the presence of strong temporal disorder, we observe that FV scaling persists, $(\alpha,\beta,z)_{\rm exp}=[0.49(1),0.23(2),1.9(2)]$, where the scaling exponents become consistent with the diffusive Edwards--Wilkinson class~\cite{Edwards1982}.
A similar trend is also reflected in the dynamics of density-density correlation functions.
In the clean limit, the correlation spreads ballistically, with the speed proportional to the Lieb-Robinson velocity. 
Under dynamic disorder, the correlations spread diffusively, yielding a diffusion constant close to the previous observation for a two-leg Bose-Hubbard system~\cite{Wienand2024}. 

\section*{Family--Vicsek scaling in the Hubbard model}\label{sec2}
In a one-dimensional system with size $L$, the FV scaling is identified using the roughness $w(L,t)$ defined as the standard deviation of a one-dimensional surface height $h(x,t)$ (Fig.~\ref{ExpProtocol}a).
The surface roughness satisfies the universal scaling, $w(L,t)=L^{\alpha}f(t/L^z)$, where the universal function $f(x)\propto x^{\beta}$ for $x\ll1$ and $f(x)\approx c$ for $x\gg1$ with some constant $c$, respectively (Fig.~\ref{ExpProtocol}b). 
The roughness exponent $\alpha$, growth exponent $\beta=\alpha/z$, and dynamic exponent $z$ determine the universality class of the system~\cite{Barabasi2009,Takeuchi2018}.
Recently, numerical studies have found the FV scaling in quantum many-body systems~\cite{Fujimoto2020,Fujimoto2021,Fujimoto2022,Bhakuni2024,Aditya2024,Moca2025}, displaying various kinds of transport from ballistic~\cite{Fujimoto2020,Fujimoto2021,Moca2025}, diffusive~\cite{Fujimoto2020,Moca2025} and super-diffusive~\cite{Aditya2024,Moca2025} ones with different sets of exponents $\alpha,\beta$, and $z$.
While most of the previous experimental studies~\cite{Erne2018,Prufer2018,Glidden2021,Huh2024,Gazo2025,Wei2022,Rosenberg2024} on the universal dynamics focused on the dynamic exponent $z$, the full validation of FV scaling requires complete scaling collapse across time and system size.

\begin{figure*}[t]
\centering
\includegraphics[width=0.9\linewidth]{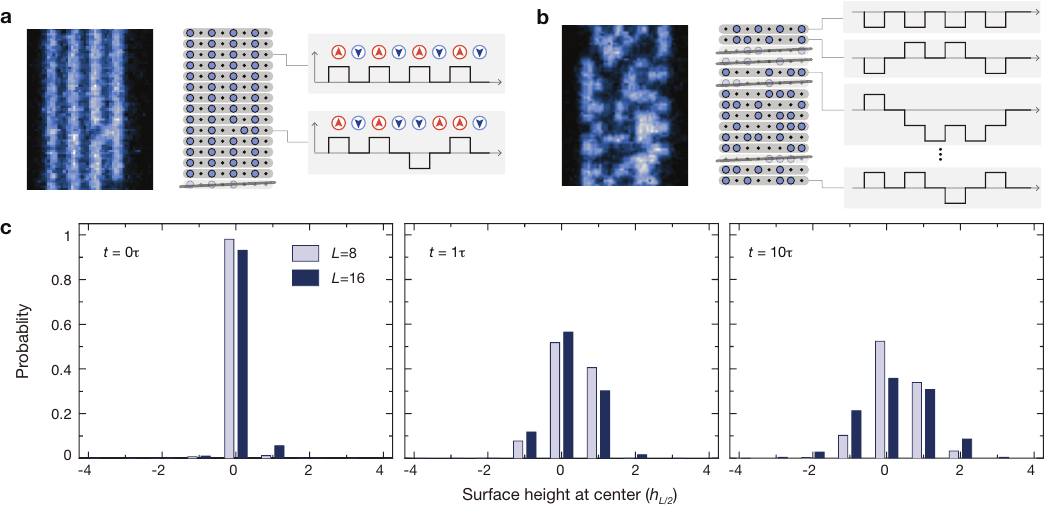}
\caption{\textbf{Quantum surface height and roughness of 1D spin chains.}
\textbf{a,} Fluorescence image of the initial CDW state and the site occupation. 
Our system of hardcore bosons is mapped to a spin-1/2 system, where the surface height is defined as the total spin of the subsystem up to site $i$. 
\textbf{b,} Atom distribution after 10$\tau$ the quench, where $\tau=\hbar/J=0.53$~ms. 
Different surface height patterns appear as a result of quantum tunneling. 
By counting the total atom number in each chain, we exclude data corresponding to imperfectly prepared initial states.
\textbf{c,} Histogram of the center surface height at $t=0,1,10\tau$ is shown for $L=8$ (light blue) and $L=16$ (dark blue).
}\label{RoughnessDef}
\end{figure*}

Using atomic quantum simulators in optical lattices, we investigate the FV scaling in the one-dimensional Bose-Hubbard model, where the Hamiltonian can be written as,
\begin{equation}
\begin{aligned}
\hat{H}_{\rm BH}&=-J\sum_{i=1}^{L-1}\left(\hat{b}_{i}^{\dagger}\hat{b}_{i+1}^{}+{\rm h.c.}\right)+\frac{U}{2}\sum_{i=1}^{L}{\hat{n}_{i}(\hat{n}_{i}-1)}.
\end{aligned}
\end{equation}
Here, $\hat{b}_i^{\dagger}$~($\hat{b}_i$) is the bosonic creation (annihilation) operator at $i$-th site, $\hat{n}_i=\hat{b}_i^{\dagger}\hat{b}_i$ is the number operator, $L$ is the length of the system, $J$ is the tunneling strength, and $U$ is the on-site interaction strength.
In a strongly interacting regime ($U/J\gg1$) at low-filling ($n_i<1$), the Bose-Hubbard Hamiltonian reduces to the spin-1/2 XX Hamiltonian,

\begin{equation}
\begin{aligned}
H_{\rm XX}&=-2J\sum_{i=1}^{L-1}{(\hat{s}_i^x\hat{s}_{i+1}^x+\hat{s}_i^y\hat{s}_{i+1}^y)},
\end{aligned}
\end{equation}
where $\hat{s}_i^{\mu}$ denotes the spin operator along $\mu\in\{x,y,z\}$ at site $i$.
Occupation of multiple atoms on a single site is suppressed by strong interactions, and the occupation number state $\ket{n_i}=\ket{0},\ket{1}$ of the Bose-Hubbard Hamiltonian corresponds to spin state $\ket{s_i^z}=\ket{\downarrow},\ket{\uparrow}$ in the XX Hamiltonian.
This model can be mapped to non-interacting spinless fermions and is integrable.

We introduce a surface height operator in the Bose-Hubbard model as~\cite{Fujimoto2020}, $\hat{h}_j(t)\equiv \sum_{i=1}^j  [ \hat{n}_i(t)-\nu  ] =\sum_{i=1}^j \hat{s}_i^z(t)$, where $\nu=\langle\hat{n}_i\rangle=1/2$ is the average site occupation.
This is motivated by the fact that in classical fluctuating hydrodynamics, density fluctuations show a similar dynamic scaling with $\partial_x h(x,t)$~\cite{Spohn2014,Das2014,Spohn2015}.
It implies the quantum ``surface height" $\hat{h}_j$ at site $j$ is constructed by integrating the local density fluctuations over space, capturing the cumulative effect of local quantum fluctuations.
The surface roughness $w(t, L)$ is then defined as the fluctuation of the surface height at the chain center $i=L/2$, $w(t, L)=[\langle\hat{h}_{L/2}^2 \rangle-\langle{\hat{h}_{L/2}} \rangle^2]^{1/2}$.
It serves as a key indicator of FV scaling~\cite{Fujimoto2021} and offers valuable insights into other important quantities such as entanglement entropy and transport of conserved quantities~\cite{Bhakuni2024}.
From the numerical study~\cite{Fujimoto2020}, this quantity exhibits the FV scaling with ballistic universality class, $[\alpha,\beta,z]_{\rm num}\simeq [0.5,0.5,1]$.

\begin{figure*}
\centering
\includegraphics[width=0.85\linewidth]{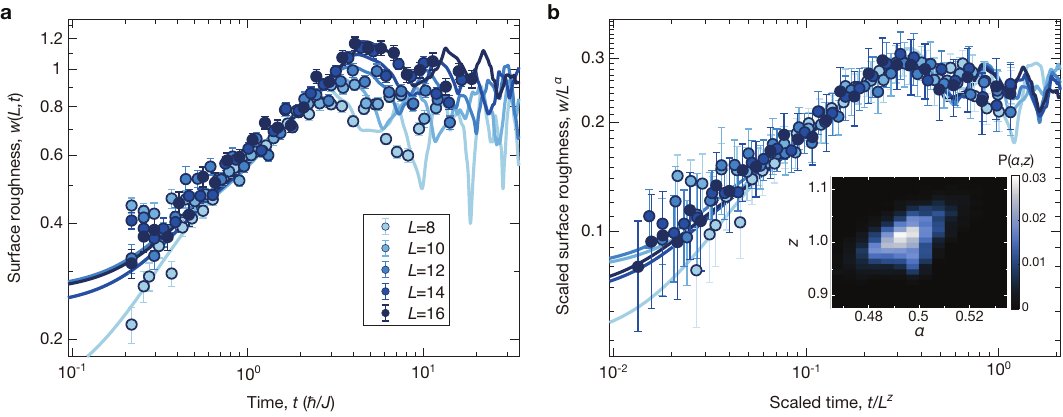}
\caption{\textbf{Ballistic quench dynamics of the XX model.} 
\textbf{a,} Surface roughness at various hold times $t\in [0,19]\tau$ for system size $L=8,10,12,14,16$. The solid lines represent numerical calculations that account for errors in the initial state.  
\textbf{b,} The curves collapse into a single curve when rescaled with the exponents $\alpha=0.49(1)$ and $z=1.00(4)$. The exponents and their errors are determined by optimizing the collapse of $10^4$ bootstrapped datasets. The joint probability of the exponents $P(\alpha,z)$ is shown in the inset.}\label{Ballistic}
\end{figure*}

We began the FV scaling experiment by preparing a CDW state of two-site periodicity as displayed in Fig.~\ref{ExpProtocol}c.
In our setup, a programmable potential created by a digital micromirror device (DMD) can be superimposed on the optical lattice, allowing us to prepare arbitrary density patterns in a two-dimensional plane~\cite{SOM}.
Dynamics are initiated by ramping the lattice depth to 10 $E_r$, where $E_r=h\times12.6$~kHz is the recoil energy of the lattice, so that $J/h=300$~Hz and $U/J=39$. 
The one-dimensional Hubbard model is realized by reducing the retroflected laser beam power in a folded-lattice geometry~\cite{SOM}.
The coupling between interlayer chains is negligible on our time scale as the intra-chain tunnelling time, $\tau=\hbar/J=0.53$~ms, is much shorter than that of the interlayer chain, $\tau_{\perp}= 300$~ms.
The total length of the one-dimensional chain $L$ is determined by imposing a potential barrier at the two ends (Fig.~\ref{ExpProtocol}d), which is essential in studying the FV scaling because it requires the finite size scaling to extract the exponents. 
Stiff wall potentials at both ends of each chain suppress hopping beyond the system size $L$, enforcing open boundary conditions.
After the quenched system undergoes a time evolution for time $t$ in a clean lattice, we freeze the hopping along the $x$-axis and take fluorescence images of the atoms to measure $L$-bit strings of the atom occupation.

\section*{Ballistic universality class}\label{sec3}

The CDW initial state has a plain surface distribution (Fig.~\ref{RoughnessDef}a) and, in an ideal case, it has no surface roughness $w(0,L)=0$. 
As time evolves, the atoms can tunnel into adjacent lattice sites, exhibiting various surface shapes (Fig.~\ref{RoughnessDef}b) and increasing the variance of the surface height (Fig.~\ref{RoughnessDef}c).  
In this regard, in a quantum system, the surface roughness grows over the hold time because of the quantum fluctuations.
To test the dynamic scaling behavior, we experimentally study the dynamics of surface roughness for various system sizes $L=8,10,12,14,16$ and plot the result in Fig.~\ref{Ballistic}a. 
Here, we post-select images satisfying $\sum_i \hat{s}_i^z=0$ because the $U(1)$ symmetry of the XX model ensures conservation of total magnetization during time 
evolution and our initial CDW state $\ket{\psi_{\rm CDW}}=\ket{\uparrow,\downarrow,\uparrow,\downarrow,...\uparrow,\downarrow}$ has zero magnetization.
For each time point, the surface roughness is computed from a bootstrapped dataset of 200 post-selected bitstrings.
In the early dynamics, the surface roughness increases as a power law, and at a later time it asymptotically saturates to a constant value (Fig.~\ref{Ballistic}a).
At longer times, we observe signatures of long-lived coherence characteristic of the integrable XX model.
That is, rather than converging smoothly to a saturated value, the surface roughness continues to fluctuate even after 10 tunnelling times, indicating the persistence of coherence.
This effect is more pronounced in smaller system sizes.

To reveal the universal scaling and extract the exponents, we rescale time and roughness by $L^z$ and $L^\alpha$, respectively, collapsing data across system sizes onto a single curve (Fig.~\ref{Ballistic}b).
The scaling exponents $(\alpha,\beta,z)$ and their uncertainties are extracted by fitting the data to the scaling form $w(t,L)=L^\alpha f(t/L^z)$, where $f(t)$ is an empirically determined universal function.
Fitting is repeated 10000 times using bootstrapped datasets, producing Gaussian-like distributions near the optimal values~\cite{Dogra2023}.
The best collapse is achieved with exponent $[\alpha,\beta,z]_{\rm exp}=[0.49(1),0.41(1),1.00(4)]$, confirming ballistic ($z=1$) dynamics and universal scaling behavior across finite system sizes.
This result can be understood from the fact that our system realizes the spin-1/2 XX model~\cite{Jepsen2020,Wei2022}, which can be mapped to a non-interacting fermion system through the Jordan-Wigner transformation.

\begin{figure*}
\centering
\includegraphics[width=0.75\linewidth]{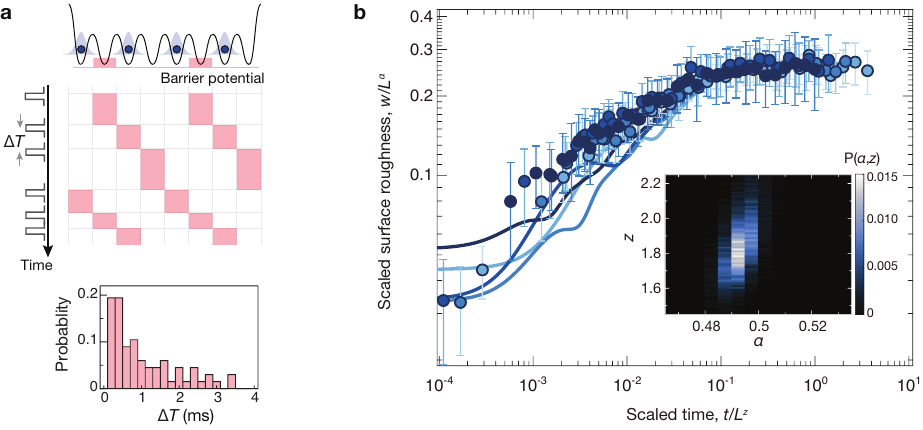}
\caption{\textbf{Diffusive quench dynamics of the XX model with dynamic disorder potentials.}
\textbf{a,} Schematic diagram of the temporal disorder. After reducing the lattice depth, the on-site potential barrier is temporarily superimposed on the lattice. 
The timing of the potential barrier is controlled by a trigger pulse (square-wave train), and the barrier potential is kept on (light-red area) for the time interval of the pulse $\Delta T$. The $\Delta T$ follows an exponential distribution (see the bottom panel).
 \textbf{b,} Scaled surface roughness at various hold times $t\in [0,380]\tau$ and different system sizes $L=12,16,20,24$ under the influence of temporal disorder potentials (see main text). 
The curves best collapse when rescaled with the exponents $\alpha=0.49(1)$ and $z=1.9(2)$. The solid lines correspond to the numerical simulations. 
The joint probability of the exponents $P(\alpha,z)$ is shown in the inset.}
\label{Diffusive}
\end{figure*}

The experimental observation can be further supported by the numerical simulation (solid lines in Fig.~3), which is obtained by solving the Schrödinger equation of the spin-1/2 XX model~\cite{SOM}.
In the simulation, we take into account the experimental imperfections of the initial state, for example, $\ket{\uparrow, \downarrow, \downarrow, \uparrow, \dots, \downarrow, \uparrow}$, which contains the correct occupation but incorrect configuration.  
We model the initial state after the post-selection as a mixed state, $\rho_0=p_0|\psi_0\rangle\langle\psi_0|+\sum_i p_i|\phi_i\rangle\langle\phi_i|$, where $\ket{\phi_i}$ represent erroneous states. 
Typically, the probability of incorrect configuration increases with system size, but the degree is less than $10\%$~\cite{SOM}. 
This initial density matrix used in theoretical calculations is chosen in accordance with the probability distribution of experimentally observed bitstrings.

\section*{Edwards--Wilkinson universality class}\label{sec4}
Having the fact that our experiments follow the FV scaling in the XX model under the static dynamics, we now turn to our second main result on the universal scaling under integrability-breaking disturbance.
In particular, we employ a temporal random potential $\epsilon_i(t)$ in each lattice site, which is given as 

\begin{equation}
\epsilon_i(t)=\Delta\sum_{n=0}^{L/4-1}\sum_{k=1}^4 \delta_{i,4n+k} \Theta_k(t),
\label{BumpEquation}
\end{equation}
where localized bump potentials with strength $\Delta=10J$ are introduced at sites $4n+k$ for nonnegative integers $n,k$.
The piecewise step function 
\begin{equation}
\Theta_k(t)=
\begin{cases}
1\quad (T_{4m+k}\le t <T_{4m+k+1})\\
0\quad (\textrm{otherwise})
\end{cases}
\end{equation}
controls the timing of these bumps (Fig.~\ref{Diffusive}a).
These local potentials induce a staggered effect, with their positions shifting cyclically at randomly determined times $t=T_{4m+k}$ for nonnegative integer $m$, starting from $T_1=0$.
The probability distribution of the time interval $\Delta T=T_{4m+k+1}-T_{4m+k}$ is set to follow an exponential distribution, 
$P[\Delta T]\propto e^{-\Delta T/\delta T}$, where $\delta T=0.53$~ms is the mean time interval. 
This distribution reflects a Poisson process and its memoryless property, where the rate of an event occurring remains constant regardless of how much time has passed.

Under the temporal disorder, we examine surface roughness dynamics for various system sizes $L=12,16,20,24$ and plot the results in Fig.~\ref{Diffusive}.  
While the surface roughness eventually saturates to the same value as in the ballistic case for a given system size, the time required to reach this saturation is significantly prolonged, where we extended the time evolution up to $t=380\tau$.
Following the same method in the XX model, we perform a finite size scaling of the dynamics and find that the surface roughness still follows the dynamic scaling, where the exponents are extracted as $(\alpha,\beta,z)_{\rm exp}=[0.49(1),0.23(2),1.9(2)]$ (Fig.~\ref{Diffusive}b), which is close to (0.5, 0.25, 2.0) of the Edwards-Wilkinson class~\cite{Edwards1982}. 
This feature is also supported in the numerical simulation, where we studied much larger system sizes up to $L=64$ and obtained the exponents as  $(\alpha,\beta,z)_{\rm sim}\simeq[0.5,0.25,2]$. 
We also study the robustness of the FV scaling under such strong random temporal kicks with different initial states and different kicks~\cite{SOM}.
All of these results give similar exponents, highlighting that the universal feature of the FV dynamic scaling is present under the temporal random potential. 

The results suggest that the surface roughness retains a universal scaling form, implying that scaling persists beyond integrability, albeit in a different universality class. The emergence of diffusive scaling in our system can be attributed to the loss of coherent motion under strong temporal disorder. 
Temporal fluctuations in the lattice potential act as a source of momentum-space dephasing, disrupting interference and destroying integrability, thereby converting the ballistic transport into an effectively stochastic diffusive process. 
Moreover, our approach is distinctive from the previous experiments~\cite{Wei2022} in that it preserves the dimensionality. 
Instead, strong temporal disorder is introduced in our experiment, from which we can isolate and probe the impact of purely time-dependent stochastic noise on the persistence and robustness of universal scaling behavior.

\begin{figure*}
\centering
\includegraphics[width=0.85\linewidth]{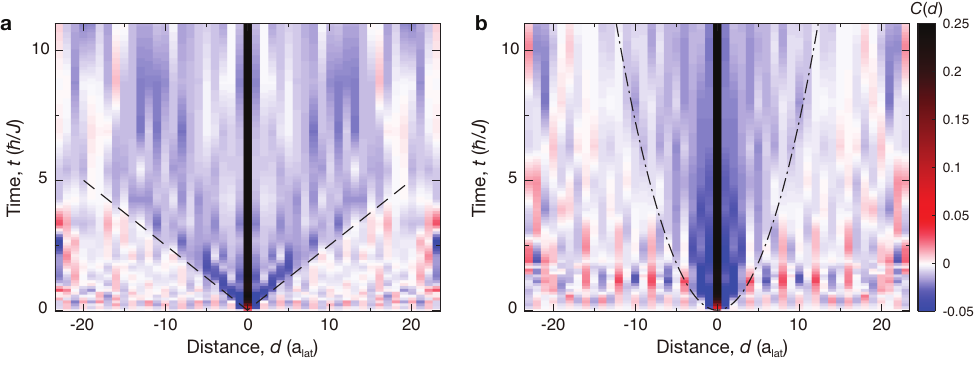}
\caption{\textbf{Dynamics of two-point correlations.} 
\textbf{a,} Time evolution of two-point correlations $C(d)$ is studied for the static XX model at system size $L=24$.  The envelope of negative correlations spreads linearly with time. The dashed line represents the ballistic expansion with the characteristic velocity $v = 4J a_{\rm lat}/\hbar$ ($a_{\rm lat}=0.752$~nm), which is derived for the equal-time spin correlation based on the Lieb-Robinson velocity $2J a_{\rm lat}/\hbar$~\cite{Bravyi2006}.
\textbf{b,} Time evolution of correlations with dynamic disorder potentials at system size $L=24$. The envelope is fit to a diffusively spreading Gaussian function $C(d)\sim \exp(-d^2/8Dt)$ where $D$ is the diffusion constant. The dashed-dot curve represents the $1/e^2$ envelope for the fitted diffusion constant $D=0.9(1)Ja_{\rm lat}^2/\hbar$.}\label{Correlation}
\end{figure*}

\section*{Dynamics of correlations}\label{sec5}

This fundamental change in the transport property can also be manifested in the correlation functions.
Using our quantum gas microscope, we measure the equal-time density–density correlation function, which can be written in terms of spin operator in the strongly interacting regime, $C(d)=\langle \hat{s}_i^z\hat{s}_j^z\rangle -\langle \hat{s}_i^z\rangle \langle \hat{s}_j^z\rangle$, where $d=|{i-j}|$ is the distance between two lattice sites $i$ and $j$.
Fig.~5 displays the dynamics of the correlation $C(d)$ for both clean and under a temporal disorder potential at a system size $L=24$. 
In the early times, negative correlations are built up over short distances because of the strong interaction between atoms or the effective exclusion principle. 
In the ballistic regime, the correlation envelope expands linearly in time, with its front propagating at the speed of $v=4Ja_{\rm lat}/\hbar$ with $a_{\rm lat}=0.752$~nm~\cite{Lieb1972,Bravyi2006,Cheneau2012}.
In contrast, under the random potential kick, the spread of negative correlations follows a square-root time dependence, consistent with diffusion.
Fitting the correlation map to a Gaussian profile with time-dependent width yields a diffusion constant of $D=0.9(1)Ja_{\rm lat}^2/\hbar$~\cite{SOM}.

\section*{Discussions}\label{sec5}
In this work, we have demonstrated the Family--Vicsek (FV) scaling in one-dimensional Bose gases in the strongly interacting regime.
By rescaling time and length with a set of exponents, the data collapse onto a single universal curve, demonstrating the universal behavior predicted by the FV scaling.
This work goes beyond most previous experimental studies of nonequilibrium quantum dynamics, which have focused on measuring a single dynamic exponent. 
Furthermore, by introducing a temporal random potential, we are able to control the universal dynamics of quantum systems from the ballistic universality class to the Edwards-Wilkinson diffusive universality class, having excellent agreement with theoretical predictions. 
Note that, while there are several definitions of the height operators as addressed in Sec. 3.3 of the supplemental material, we numerically confirm that the choice of the height operators does not affect the universal roughness growth.

These findings suggest many intriguing possibilities for future investigations.
One particularly interesting question is the applicability of the macroscopic fluctuation theory in quantum many-body systems.
The macroscopic fluctuation theory is a modern framework in classical systems that describes the hydrodynamics with large-scale fluctuations~\cite{Bertini2015}.
In one-dimensional stochastic models, the symmetric simple exclusion process could lead to a diffusive scaling~\cite{Kipnis1986}, whereas in the asymmetric exclusion process, their dynamics can follow the KPZ universality class~\cite{Tracy2009}.
Recently, a quantum analogue of symmetric (asymmetric) simple exclusion process has been studied in a one-dimensional spin model~\cite{Bernard2019,Bauer2019,Jin2020}, which can be experimentally realized by elaborating the techniques used in this experiment.
Another immediate future work is to explore other types of universal dynamics besides the conventional universality class.
For example, the FV scaling in the random disorder potentials may exhibit anomalous scaling in the delocalized regime~\cite{Fujimoto2021}, and rich dynamics like subdiffusive, ballistic, and superdiffusive behavior have been predicted in a driven quasi-periodic disorder system~\cite{Aditya2024}. 
Finally, the finite-size scaling method employed in this work is suited for extracting the critical exponent of unconventional phase transitions both in and out-of-equilibrium quantum many-body systems, such as non-Hermitian systems~\cite{Wei2017} and disordered many-body systems~\cite{Abanin2019}.

\begin{acknowledgments}
We acknowledge discussions with Tomohiro Sasamoto, Björn Sbierski, and Soonwon Choi.
Y.K. was supported by JSPS KAKENHI Grant No. JP24K00557. R.H. was supported by JST ERATO Grant Number JPMJER2302, and by JSPS KAKENHI Grant No. JP24K16982. K.F. was supported by JSPS KAKENHI Grant No. JP23K13029. 
J.-y.C. is supported by the National Research Foundation of Korea (NRF) Grant under Project No. RS-2023-00256050, RS-2023-NR119928, 2023M3K5A1094812, and RS-2025-02220735.  

\end{acknowledgments}

\bibliography{FVscaling,maxbibnames=5}

\newcommand{\beginsupplement}{%
        \setcounter{table}{0}
        \renewcommand{\thetable}{S\arabic{table}}%
        \setcounter{figure}{0}
        \renewcommand{\thefigure}{S\arabic{figure}}%
        \setcounter{equation}{0}
        \renewcommand{\theequation}{S\arabic{equation}}%
     }
\clearpage
          
\newpage
\beginsupplement

\onecolumngrid 
\baselineskip18pt

\begin{center}
\large{\textbf{Supplemental Materials for \\ ``Universal Family--Vicsek scaling in quantum gases far from equilibrium"}}\\
\end{center}

\section{Experimental systems}\label{sys}
\subsection{Single-site addressing system.} 
Applying a programmable optical potential with single-site resolution in the two-dimensional square lattice is the key ingredient in our experiments. 
The programmable potential is generated by using a digital micromirror device (DMD) (Vialux V-7001), which reflects an incident light into programmable patterns using small mirrors that are in either the `on' or `off' state.
We use a blue-detuned incoherent light at 650~nm generated from a superluminescent diode (EXS210030-03) that is amplified fourfold using two double-pass tapered amplifier configurations to achieve a maximum power of 50~mW at the fiber output \cite{Bolpasi2010,Hur2025}.
The DMD is located at the image plane of the magnification of a 150$\times$ imaging system, which projects the potential pattern from the DMD onto the atomic plane~\cite{Fukuhara2013}.  
The large magnification is realized in two stages: a 4$\times$ imaging system for the DMD is integrated into the 37.5$\times$ microscopy system, which is used for imaging atoms with single-site resolution~\cite{Kwon2022}. 
Consequently, one lattice site corresponds to $8.25\times8.25$ pixels of the DMD.
This enables us to realize grayscale patterns by dithering the pattern into binary images using an error diffusion algorithm~\cite{Dorrer2007}. 

The affine transformation matrix that maps the pattern on the DMD to the atom plane is calibrated by imposing a stripe pattern on a unity-filled Mott insulator.
We scan the pitch, rotation angle, and phase of the stripe pattern such that the repulsive potentials are injected to every other lattice site, creating a stripe pattern on the Mott insulator~\cite{Choi2016,Rubio2019}.
The optimal pitch was 16.5 pixels along both axes, corresponding to two lattice sites, and the optimal rotation angle was $7.5^\circ$. 
While the phase of the stripe patterns of the DMD was negligible, the phase of the optical lattice in the atom plane had a slow phase drift on the order of 0.5 lattice sites/hour.
The relative phase difference between the DMD pattern and the lattice is stabilized by monitoring the phase of the 2D lattice from the fluorescence images, which is sent to adjust the phase of the CDW image. 
This feedback is performed at the end of each 40 seconds of a single experimental run, and there are very few cases where abrupt phase jumps occur and the lock is broken. 
This can be confirmed by analyzing the initial state CDW fidelity.

\begin{figure*}
\centering
\includegraphics[width=0.8\linewidth]{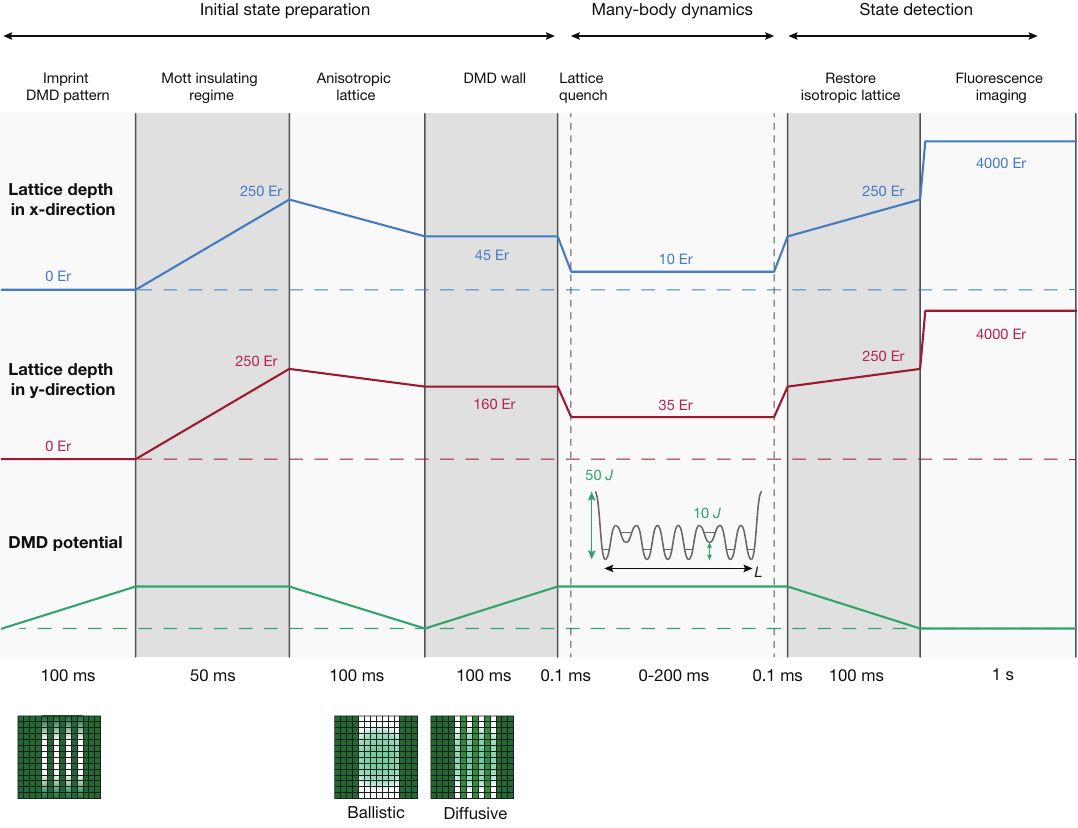}
\caption{
\textbf{Schematic diagram of experimental sequence.} The experiments are composed of three main subsections: initial state preparation, many-body dynamics, and detection. 
The main tuning parameters in the experiments are lattice depth along the $x$ and $y$ directions, laser intensity for DMD, and patterns displayed on the DMD.  
Axes are not drawn to scale.
Images on the top panel display the patterns on the DMD, where each box represents the lattice sites, and the addressed sites are marked by the gray marks. 
The first image is a stripe patterns with side walls that keeps the chain length to $L$.
A weak gradient along the $y$ axis provides a weak confinement effect in the vertical axis, enhancing the fidelity of the CDW state. 
After preparing the CDW state, we change the images to either clean for ballistic transport or temporal disorder for diffusive transport. 
The barrier height is about $h\times15$~kHz$=50J$, and the local on-site potential for the temporal disorder is $h\times3$~kHz$=10J$.
The parabolic shading in the second image is intended to compensate for harmonic confinement from the lattice beam.
}
\label{Sequence}
\end{figure*}

\subsection{Initial state preparation.} 
We follow a standard sequence for creating a unity-filling Mott insulating state~\cite{Hur2024}.  
Degenerate Bose gases of $^{7}$Li atoms are prepared in a single plane of vertical optical lattice.
Increasing the lattice potential in a two-dimensional plane to an atomic limit, a low-entropy Mott insulating state can be generated with a filling of 0.96(1).
An experimental sequence for preparing the charge density wave state is shown in Fig.~\ref{Sequence}.
Before the lattice ramp, we introduce high potential barriers at target lattice sites, which in turn can create an arbitrary density pattern when the chemical potential of the system is smaller than the onsite interaction energy $U$ in the atomic limit.
For example, we use a stripe pattern aligned along the $y$-axis on the DMD to generate a charge density wave state along the $x$-axis.
The length of the CDW state $L$ is determined by placing a potential barrier at both ends, and we properly control the atom number to avoid $n=2$ filling after the lattice ramp.  
The potential height of the addressed lattice site is about $\Delta \epsilon_i/h \approx 10-15~\rm{kHz}$.
To maximize the fidelity of the state of interest, we add entropy reservoir sites that have slightly higher potential than our region of interest.
Most of the entropy of the system is then distributed at the reservoir sites while leaving the region of interest at a lower entropy state.

Then, we adiabatically reduce the power in two of the lattice arms by rotating a quarter-waveplate mounted on a motorized stage in front of the retro-reflecting mirror.
This transforms the two-dimensional lattice to an anisotropic lattice $V_x=45E_r, V_y=158E_r$, where $E_r$ is the recoil energy of the optical lattice.
At the same time, we reduce the beam power for the DMD and change the DMD pattern to either clean with both ends or a local bump potential. 
While we change the pattern, tunneling is still negligible in both directions, and the atoms are well localized in each lattice site.
Once the patterns have been changed, we increase the DMD laser beam and suddenly lower the lattice depths to $V_x=10E_r, V_y = 35E_r$ in 0.1ms, respectively. 
The lattice quench initiates the intra-chain dynamics along the $ x$-direction while the inter-chain dynamics remain suppressed along the $ y$-direction, such that the dynamics of the system are governed by the 1D Bose-Hubbard Hamiltonian.
After a sufficient amount of hold time, we freeze the dynamics by raising the lattice depths, restoring an isotropic lattice, and taking fluorescence images at maximum lattice depth.\\

\subsection{System calibration.} 
\begin{figure*}[t]
\centering
\includegraphics[width=0.9\linewidth]{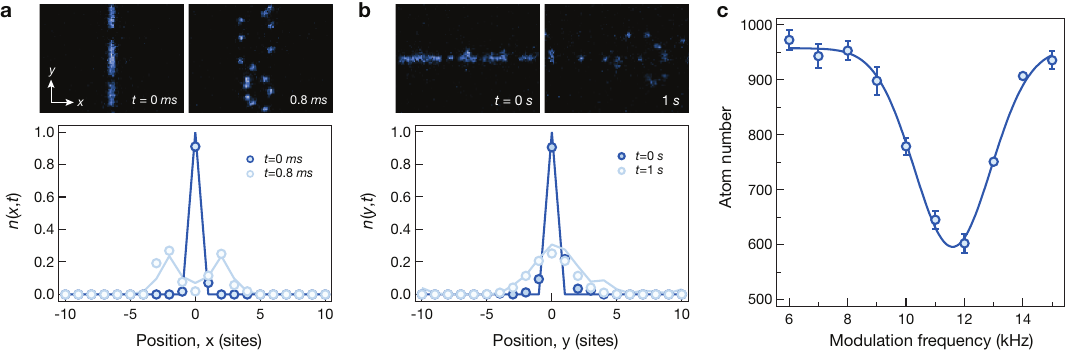}
\caption{
\textbf{Hamiltonian parameter calibration.}
\textbf{a,} Intra-chain coupling $J$ is measured by preparing single atoms in each chain and observing a one-dimensional quantum walk.
\textbf{b,} Inter-chain coupling $J_\perp$ is similarly measured by preparing atoms in a single chain and observing the spread to other chains. 
\textbf{c,} The on-site interaction $U$ is measured by preparing a unity filling Mott insulator and modulating the lattice depth.
Doublon-hole formation is enhanced when the modulation frequency is resonant to $U$, which is observed as a decrease in single occupancies.
}
\label{Cal_UJ}
\end{figure*}

The tunneling strength $J$ is calibrated from the one-dimensional quantum walk of single atoms on the optical lattice \cite{Preiss2015}.
To measure intra-chain tunneling along $x$, we initialize a line of atoms along the $y$-axis, representing ensembles of single atoms with the same initial position.
By performing the same quench sequence in our experiments, we obtain the inter-chain tunneling $J$ from the one-dimensional quantum walk along $x$ (Fig.~\ref{Cal_UJ}a). 
Fitting the density profile to the analytic solution 
\begin{equation}
n(x,t)=|\mathcal{J}_x(2Jt/\hbar)|^2
\end{equation}
where $\mathcal{J_\nu}$ is the $\nu$th order Bessel function of the first kind, we measure a tunneling rate of $J/h=300(20)\rm{Hz}$.

Inter-chain tunneling along $y$ is similarly measured by preparing a line of atoms along the $x$-axis.
In this case, instead of a one-dimensional quantum walk, the dynamics becomes two-dimensional, and tunneling along $x$ scrambles phase information, leading to an effectively classical random walk behavior along $y$ (Fig.~\ref{Cal_UJ}b).
The density profile along $y$ is thus given as a normal distribution, from which we extract
$J_\perp/h=0.2(1)\rm{Hz}$.

\begin{figure*}
\centering
\includegraphics[width=0.9\linewidth]{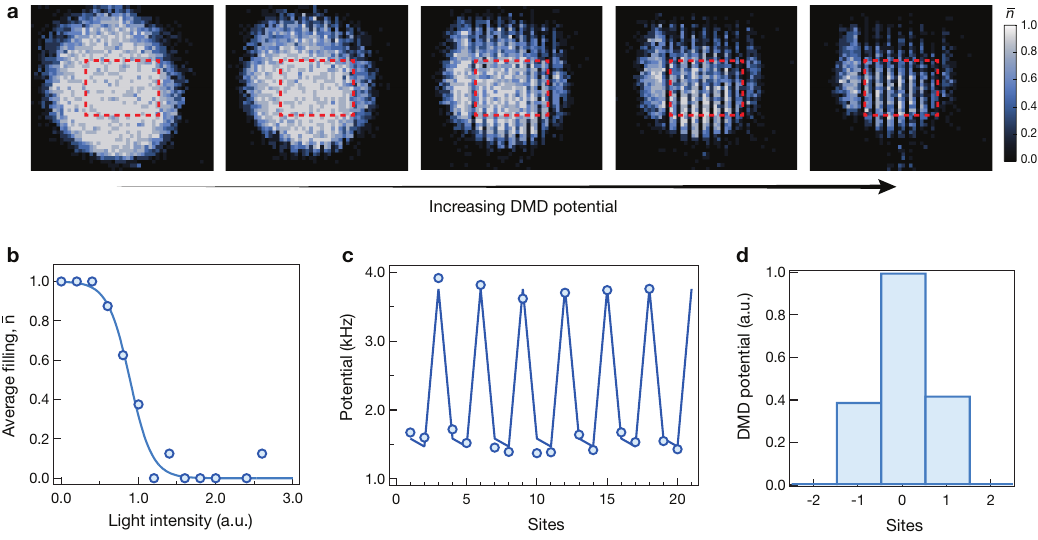}
\caption{
\textbf{DMD potential calibration.}
\textbf{a,} The response of the atoms to the increasing DMD potential. As the potential of the DMD increases, the filling of atoms at each site is reduced because of the repulsive potential of the DMD. The filling of the atoms ($\bar{n}$) in the central region (dotted red box) is measured to calibrate the applied DMD potential in terms of the injected light intensity. The overall harmonic curvature can also be measured with the averaged atom density without the applied DMD potential.
\textbf{b,} Averaged filling of atoms at each site in terms of the applied light intensity. The curve follows a sigmoid-like function when the atom is sufficiently cold. The chemical potential became zero when the average filling is $\bar{n} = 0.5$. Thus, we can calculate the potential of the DMD light at each site from the knowledge of the Harmonic curvature. 
\textbf{c,} The calibrated potential with averaging in the vertical axis. The measured DMD potential well matches the calculated potential obtained by considering the broadening effects of the point spread function of the imaging system.
\textbf{d,}The broadening of the single-line DMD potential. Because of the limited optical resolution, the single line potential affects the neighboring sites. We measured the effect of the single-line potential on the neighboring sites by finding the convolution matrix that describes the measured DMD potential most. The potential of the neighboring sites is about 0.4 of the peak potential. 
}
\label{Cal_DMD}
\end{figure*}

We measure the on-site interaction $U$ by lattice modulation spectroscopy \cite{Stoferle2004}.
A unity-filled Mott insulator with 1000 atoms is prepared at depths $V_x=10E_r, V_y=35E_r$.
We then modulate the lattice depth by $2E_r$ and scan the modulation frequency $\omega$.
Resonant light-assisted tunneling occurs when $\hbar\omega\simeq U$, leading to enhanced doublon population.
This resonance is observed as a dip in atom number with parity projected measurements as shown in Fig.~\ref{Cal_UJ}c, from which we measure $U/h=11.6(2)~\rm{kHz}$.
The large ratio of $U/J=39$ verifies that our system is indeed within the hardcore boson limit.

Calibrating the repulsive on-site potential generated by DMD patterns is crucial for preparing high-fidelity initial states and accurately engineering the on-site potential terms in the Hamiltonian.
The site-resolved on-site potential $\varepsilon_i$ is calibrated by measuring the atomic density of a Mott insulator while varying the DMD light intensity (Fig.~\ref{Cal_DMD}a and b).
At low filling, where each site is occupied by at most one atom, the local density follows the distribution
\begin{equation}
    \bar{n}_i=\frac{1}{1+\exp[\beta(\varepsilon_i-\mu)]}
    \label{boseeinstein}
\end{equation}
where $\beta=(k_B T)^{-1}$ and $\mu$ is the global chemical potential.
We first fit the density profile of the Mott insulator without any applied DMD potential to extract $\beta$, $\mu$, and the harmonic trap curvature in units of $U$.
Then, by scanning the input laser intensity to the DMD and fitting the resulting densities to (Fig.~\ref{boseeinstein}), we reconstruct the site-resolved potential landscape created by the DMD.
Fig.~\ref{Cal_DMD}c shows the reconstructed potential map of a DMD pattern designed to create a delta-like potential every three lattice sites.
Due to the finite point spread function (PSF) of the imaging system and the finite size of the DMD pixels, the potential extends beyond the targeted sites.
From the measured potential map, we quantify that approximately 40\% of the peak amplitude leaks to nearest-neighboring sites, while next-nearest-neighbor crosstalk remains negligible ($<1\%$).

\begin{figure}[t]
\includegraphics[width=0.85\linewidth]{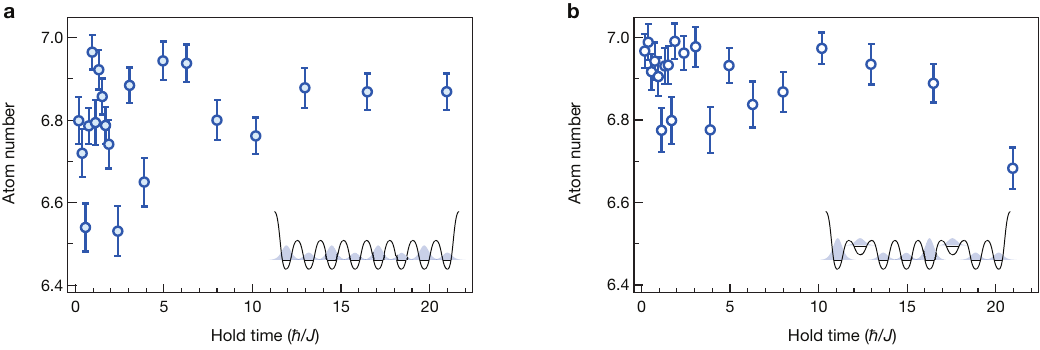}
\caption{
\textbf{Atom number conservation.}
 The total atom number for ensembles with a system size of $L=16$ is averaged without post-selection. No significant atom loss is observed for both in \textbf{a,} clean and \textbf{b,} perturbed lattices.
}
\label{atomN}
\end{figure}

\section{Data Analysis}
\subsection{Post-selection by atom number} 
\begin{figure}[b]
\centering
\includegraphics[width=0.9\linewidth]{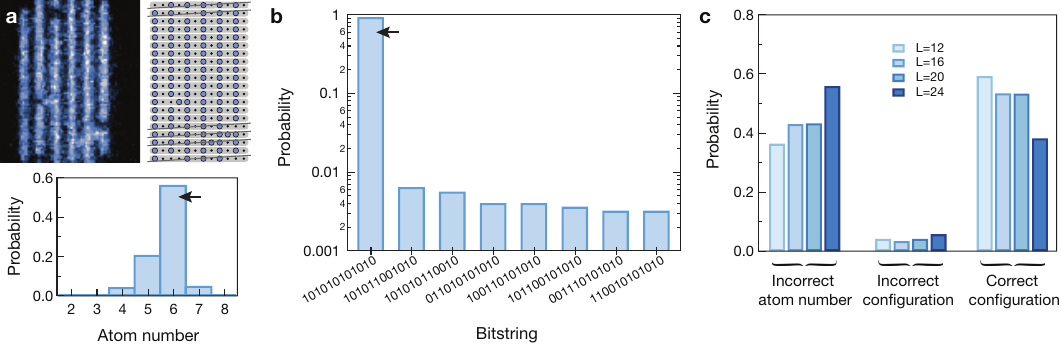}
\caption{
\textbf{Statistics of initial state.}
\textbf{a,} From a fluorescence image (upper left), we reconstruct the site occupation number on each chain (upper right). 
The histogram of atom number in the initial state is shown for a system size of $L=12$ (lower panel).
We post-select outcomes that only have the exact target atom number $L/2=6$, marked by a black arrow in the histogram.
\textbf{b,} Probability distribution of various bitstrings for a given atom number, $N=6$. An ideal initial state is prepared with a fidelity of 93\% (black arrow). 
\textbf{c,} Occurrence probability for different system sizes used in the experiment.
Most of the errors have incorrect atom numbers, which we mitigate by post-selecting the outcomes based on the atom number.
}
\label{PostSelect}
\end{figure}

To mitigate errors in our initial state preparation, we post-select the ensembles based on the total atom number on each chain (Fig.~\ref{PostSelect}a).
With such a post-selection protocol, we are able to remove the erroneous states with incorrect numbers, reducing the Hilbert space to the span of states that have the correct atom number.
We confirm that the average number of atoms without post-selection is well conserved within the time window of our observation, verifying the number conservation of our Hamiltonian and validating our post-selection method (Fig.~\ref{atomN}).
The fidelity of a $L=12$ chain reaches up to 93\% after post-selection, and the most probable erroneous states are shown in Fig.~\ref{PostSelect}b.
The initial state fidelity naturally decreases with system size. 
Without post-selection, the fidelity drops from 60\% at $L=12$ to 38\% at $L=24$, whereas with post-selection, it is better preserved from 93\% at $L=12$ to 87\% at $L=24$.

\subsection{Determination of diffusion constant} 
The propagator $G(x,t)$ for a single particle in one-dimensional diffusive motion is given as a Gaussian with a spread of $\sigma^2=2Dt$.
\begin{equation}
    G(x,t)=\frac{1}{\sqrt{4\pi Dt}}\exp({-\frac{x^2}{4Dt}})
\end{equation}
The equal-time correlation $C(d)$ is thus proportional to the spatial convolution of the propagators and spreads with a Gaussian width of $\sigma_C^2 =2\sigma^2=4Dt$.
\begin{equation}
    C(d)\sim G(x,t)*G(x,t)=\frac{A}{\sqrt{16\pi Dt}} \exp({-\frac{x^2}{8Dt}})
\end{equation}
We measure the diffusion constant of our system by fitting the spread of negative correlations in $t\in[0,10] \hbar/J$ with two fitting parameters $A$ and $D$, which yields $A=0.42(3)$ and $D=0.9(1)Ja_{\rm lat}^2/\hbar$.
The diffusion constant is close to that for the experimental measurement and theoretical predictions of a two-leg ladder system, where inter-chain coupling introduces diffusive transport to the system \cite{Wienand2024}.

\section{Theoretical methods}
\subsection{Detail of the numerical simulations}
We shall describe the setup and the method for our numerical simulations. Our simulations utilize the fact that the time-evolved density matrix remains a Gaussian state under the unitary time evolution, and thus, we can efficiently calculate the roughness in quantum many-body dynamics. 

We define our theoretical model and introduce the physical quantity of our interest, namely the roughness. 
The experiment realizes the one-dimensional Bose-Hubbard model with the strongly repulsive interaction, and thus the effective theoretical model becomes the spin-1/2 XX model as explained in the main text. By applying the Jordan-Wigner transformation to the quantum spin model, we find that the Hamiltonian is given by non-interacting spinless fermions. Hence, in the following, we focus on the Hamiltonian for non-interacting fermions on a one-dimensional lattice. 

Suppose that we have a one-dimensional lattice labeled by $\Lambda \coloneqq \{1,2,...,L \}$, where $L$ is a multiple of four. We denote the annihilation and creation operators at site $j \in \Lambda$ for the spinless fermions by $\hat{a}_j$ and $\hat{a}_j^{\dagger}$, respectively. Then, the Hamiltonian is defined by 
\begin{align} 
\hat{H} \coloneqq - J\sum_{j=1}^{L-1} \left( \hat{a}^{\dagger}_{j+1} \hat{a}_{j} + \hat{a}^{\dagger}_{j} \hat{a}_{j+1} \right) + \sum_{j=1}^{L} \epsilon_j(t) \hat{a}^{\dagger}_j \hat{a}_j.
\end{align}
Here, $\epsilon_j(t)$ is a spatially and temporally varying potential, and we set the time $\hbar/J$ characteristic of the hopping to unity, where $J$ is the hopping parameter of the Bose-Hubbard model. This model is equivalent to the one-dimensional Bose-Hubbard model with the hard-core interaction up to some constants. We denote the quantum state at time $t$ by $\ket{\psi(t)}$, which evolves according to the Schrödinger equation, $ {\rm i} d \ket{\psi(t)} /dt = \hat{H} \ket{\psi(t)}$. If the initial state is mixed, we denote the density matrix at time $t$ by $\hat{\rho}(t)$ and assume it to obey ${\rm i} d \hat{\rho}(t)/dt = [ \hat{H}, \hat{\rho}(t) ]$. For a pure state, we sometimes write $\hat{\rho}(t) = \ket{\psi(t)} \bra{{\psi(t)}}$.

In this work, we use three kinds of potential $\epsilon_{j}(t)$. For this purpose, we first define a time sequence $\{T_{0} \coloneqq 0, T_1, T_2, .... \}$, which is generated by the exponential distribution for $T_{m+1} - T_{m}$ defined by 
\begin{align} 
{\mathbb P} \left[ T_{m+1} - T_{m} = \tau \right] \coloneqq \dfrac{1}{T_{\rm exp}} \exp(- \tau/T_{\rm exp})
\end{align}
with an experimental parameter $T_{\rm exp}$.
Under this setup, we consider the following potentials:
\begin{align}
&\text{(V1) zero potential} \nonumber\\
&\quad\quad \epsilon_j(t) =0, \\
\nonumber\\
&\text{(V2) 4-site-bump potential} \nonumber\\
&\quad\quad \epsilon_j(t) = \Delta \sum_{m=0}^{\infty}\sum_{\alpha=0}^3\sum_{k=0}^{L/4-1} \delta_{j, 4k+1+\alpha} \Theta(t-T_m)\Theta(T_{m+1}-t) \chi\left[ m \equiv \alpha \:({\rm mod} 4) \right], \\
\nonumber\\
&\text{(V3) random potential} \nonumber \\
&\quad\quad \epsilon_j(t) = \Delta \sum_{m=0}^{\infty} x_{j} \Theta(t-T_m)\Theta(T_{m+1}-t) \chi\left[ m \equiv 0 \:({\rm mod} 2) \right].
\end{align}
Here, $\Delta$ is a potential strength and $x_{j}$ is a random variable generated by a uniform distribution with the range $[0,1]$. The function $\chi[\bullet]$ is the indicator function, equal to 1 when the condition holds and 0 otherwise. Note that $x_{j}$ is sampled for each site $j$ independently. We also note that the random potential is present only for even $m$ for (V3). As shown later, the roughness dynamics with (V1) belong to the ballistic class, while those with (V2) or (V3) belong to the Edwards-Wilkinson class.  

As to the initial states, we consider the following four different states: 
\begin{align}
&\text{(I1) experimental initial state} \quad 
\hat{\rho}(0) = \sum_{n=1}^{N_{\rm exp}} p_n \ket{n}\bra{n}, \label{eq:initial1} \\
&\text{(I2) 2-site-periodic initial state}  \quad
\ket{\psi(0)} = \prod_{k=1}^{L/2} \hat{a}_{2k-1} ^{\dagger} \ket{0}, \label{eq:initial2} \\
&\text{(I3) 3-site-periodic initial state}  \quad
\ket{\psi(0)} = \prod_{k=1}^{\lfloor L/3 \rfloor} \hat{a}_{3k-2}^{\dagger} \hat{a}_{3k-1}^{\dagger} \times \prod_{j=1}^{L {\rm mod} 3} \hat{a}_{3\lfloor L/3 \rfloor + j}^{\dagger} \ket{0}, \label{eq:initial3}\\
&\text{(I4) 4-site-periodic initial state}  \quad
\ket{\psi(0)} = \prod_{k=1}^{L/4} \hat{a}_{4k-3}^{\dagger} \hat{a}_{4k-2}^{\dagger} \ket{0}, \label{eq:initial4}
\end{align}
where $\ket{0}$ denotes the vacuum. The state $\ket{n}$ represents a Fock state and $p_n$ is a probability for finding the state $\ket{n}$, which is experimentally determined under the assumption that the initial density matrix is separable. The number of the Fock states observed experimentally is denoted as $N_{\rm exp}$.

The physical quantity of our interest is roughness. 
The height operator is defined as
\begin{align}
\hat{h} \coloneqq \sum_{j=1}^{L/2} \left( \hat{a}^{\dagger}_j \hat{a}_j -\nu \right)
\end{align}
with the filling factor $\nu$.
Then, the roughness $w(L,t)$ is defined by
\begin{align} 
w(L,t) \coloneqq \sqrt{ \overline{ {\rm Tr}[ \hat{\rho}(t) \hat{h}^2 ]   } - \left( ~ \overline{ {\rm Tr}[ \hat{\rho}(t) \hat{h}]   } ~ \right)^2 }. \label{eq:w}
\end{align}
Here, the overline $\overline{\bullet}$ denotes the ensemble average over the random-kick-time sequence $\{ T_m \}_{m \in \mathbb{Z}_{>0}}$ and the random variables $x_{j}$.

We calculate the roughness $w(L,t)$ from a density matrix evolving under the von Neumann equation. For this purpose, we utilize the fact that all initial states are Gaussian states or sums of Gaussian states, for which the Wick theorem holds in the dynamics. Thus, it is sufficient to calculate the two-point correlation function $C_{j,k}(t) \coloneqq {\rm Tr} \left[  \hat{\rho}(t) \hat{a}_{j}^{\dagger} \hat{a}_{k} \right]$. 
In our model, we can derive the equation of motion for $C_{j,k}(t)$ as
\begin{align} 
{\rm i}\dfrac{d}{dt} C_{j,k}(t) = C_{j+1,k}(t) + C_{j-1,k}(t) - C_{j,k+1}(t) - C_{j,k-1}(t) + \left( \epsilon_{k}(t) - \epsilon_{j}(t) \right) C_{j,k}(t). 
\end{align}
We numerically solve this equation by the 4th-order Runge-Kutta method, obtaining the time evolution of the roughness $w(L,t)$. For example, we can express the following quantities by $C_{j,k}(t)$:
\begin{align} 
{\rm Tr}[ \hat{\rho}(t) \hat{h}^2 ] - \left({\rm Tr}[ \hat{\rho}(t) \hat{h}] \right)^2&=  \sum_{j=1}^{L/2} C_{j,j}(t)-  \sum_{j=1}^{L/2}\sum_{k=1}^{L/2} |{C_{j,k}(t)}|^2,
\end{align}
which can be derived using the Wick theorem. 
Thus, we can numerically calculate the roughness $w(L,t)$.

\begin{figure}
\begin{center}
\includegraphics[keepaspectratio, width=0.95\linewidth]{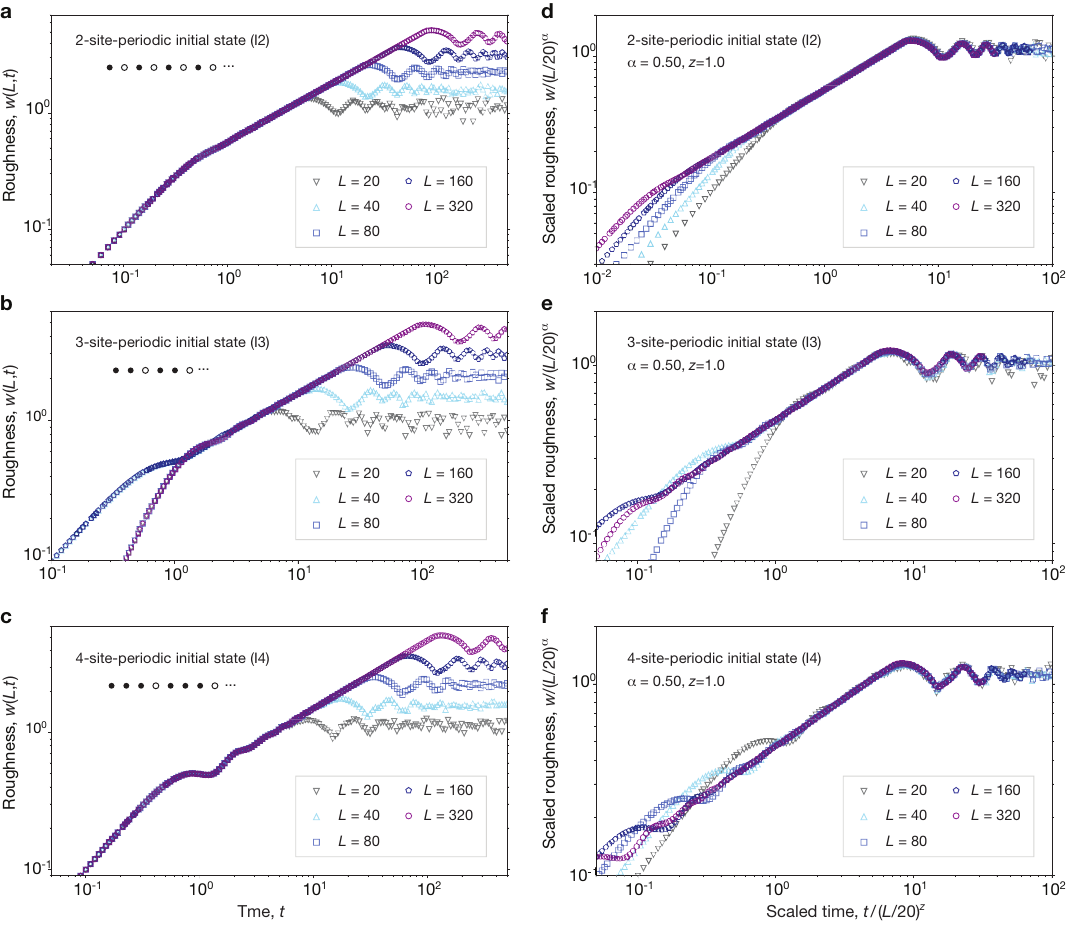}
\caption{
\textbf{Numerical results for the roughness growth in a clean system.} 
\textbf{a-c,} the panels show the time dependence of $w(L,t)$ with (I2) 2-site-periodic initial state, (I3) 3-site-periodic initial state, and (I4) 4-site-periodic initial state, respectively.  \textbf{d-f,} Rescale the abscissa and ordinate of (a-c) by $(L/20)^z$ and $(L/20)^{\alpha}$ with the scaling exponents $\alpha$ and $z$. 
The exponents in each panel are $(\alpha, z) = (0.50, 1.0)$ for all the figures.
} 
\label{N_fig1} 
\end{center}
\end{figure}

\subsection{Numerical results for the robustness of the observed Family-Vicsek scaling}
In the main text, we show only the results of the roughness dynamics with the fixed initial states (I1) and the specific potential (V1) or (V2). Here, we shall numerically discuss the robustness of the Family-Vicsek scaling reported in the main text against choices of the initial states and the kicked potentials. In what follows, we shall provide numerical evidence supporting the experimental results, which observed the universal nature of roughness growth in the quantum many-body dynamics. 

Regarding the ballistic class, we numerically study how robust the ballistic class is against various initial states. Our numerical simulations here focus on the case with (V1) zero potential $\epsilon_j(t)=0$ and the three different initial states [see (I2)-(I4)]. Fig.~\ref{N_fig1} displays the numerical results for the time evolution of the roughness. They clearly show that the ballistic class is insensitive to the different initial states, as the scaling exponents are not altered at all.  

\begin{figure}
\begin{center}
\includegraphics[width=0.95\linewidth]{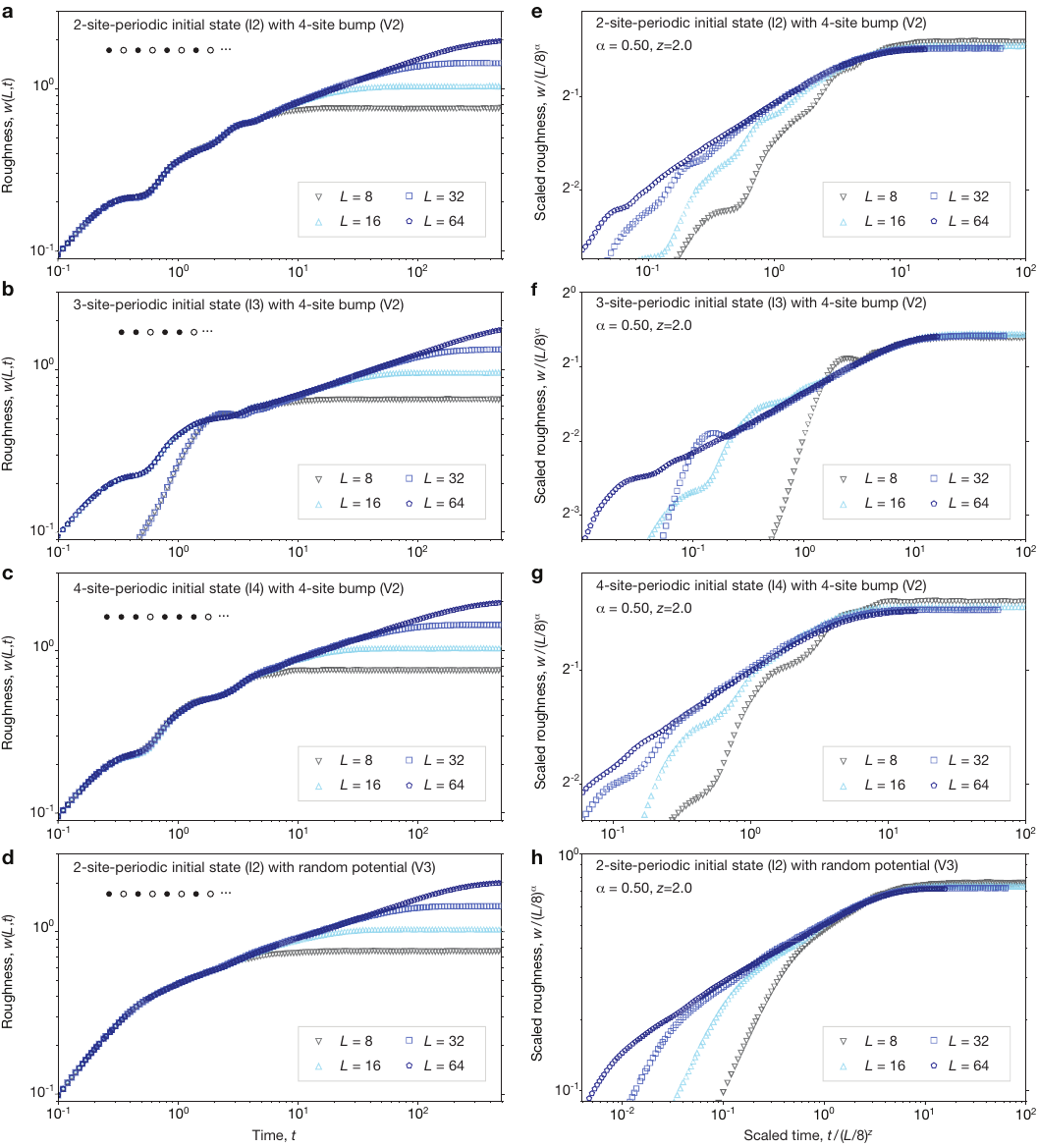}
\caption{
\textbf{Numerical results for the roughness growth under time-dependent potentials.} 
\textbf{a--c,} Time dependence of $w(L,t)$ for 4-site-bump potential (V2) with (I2) 2-site-periodic initial state, (I3) 3-site-periodic initial state, and (I4) 4-site-periodic initial state, respectively. 
\textbf{d,} Time dependence of $w(L,t)$ for (V3) with (I2) 2-site-periodic initial state. 
\textbf{e--h,} In the panels, we respectively rescale the abscissa and ordinate of panels (a-d) by $(L/8)^z$ and $(L/8)^{\alpha}$ with the scaling exponents $\alpha$ and $z$. 
The exponents in each panel are $(\alpha, z) = (0.50, 2.0)$. 
} 
\label{N_fig2} 
\end{center}
\end{figure}

For the Edwards-Wilkinson class, we numerically study the influence of the initial state and the kicked potential on the scaling exponents. The parameters used here are $T_{\rm exp} = 1$ and $\Delta=10$. Fig.~\ref{N_fig2} (a--c) and (e--g) show the roughness dynamics with (V2) 4-site-bump potential starting from the three different initial states [see (I2)-(I4)]. On the other hand, Fig.~\ref{N_fig2}d and h show the dynamics with (V3) random potential starting from (I2) 2-site-periodic initial state. All these numerical results demonstrate that the scaling exponents are very close to the exponents of the Edwards-Wilkinson class.

\subsection{Numerical results for the different definitions of the roughness}
There are several ways to define the height operator in quantum many-body systems. The following is the height operator $\hat{h}_j~(j=1,2)$ and the roughness $w_j~(j=1,2)$ used in previous literature:
\begin{align} 
&\textrm{(i)~Refs.~\cite{Jin2020,Fujimoto2020,Fujimoto2021,Fujimoto2022,Balducci2022,Balducci2023,Cecile2023,Aditya2024,Bhakuni2024}} \nonumber\\
&\quad\quad\quad~ \hat{h}_1(t) \coloneqq \sum_{j=1}^{L/2}  \left( \hat{n}_{j}(t) - \nu \right),  \\
&\quad\quad w_1(L,t) \coloneqq \sqrt{ {\rm Tr} \left(\hat{\rho}(0) \hat{h}_1(t)^2 \right) - \left[ {\rm Tr} \left(\hat{\rho}(0) \hat{h}_1(t) \right) \right]^2}, \label{eq:h1}\\
\nonumber\\
&\textrm{(ii)~Refs.~\cite{McCulloch2023,Moca2025}} \nonumber\\
&\quad\quad\quad~ \hat{h}_2(t) \coloneqq \sum_{j=1}^{L/2}  \left( \hat{n}_{j}(t) - \hat{n}_{j}(0) \right) = \hat{h}_1(t) - \hat{h}_1(0),  \\
&\quad\quad w_2(L,t) \coloneqq \sqrt{ {\rm Tr} \left(\hat{\rho}(0) \mathcal{T} \hat{h}_2(t)^2 \right) - \left[ {\rm Tr} \left(\hat{\rho}(0) \hat{h}_2(t) \right) \right]^2}, \label{eq:h2}
\end{align}
where $\hat{n}_{j}(t) \coloneqq \hat{a}^{\dagger}_j(t) \hat{a}_j(t)$ and $\mathcal{T}$ are the particle-number operator at site $j$ and the time ordering operator, respectively. The time dependence of the operators is given by the Heisenberg description under a given Hamiltonian. In Ref.~\cite{McCulloch2023}, $\hat{h}_2$ is called a charge transfer [see Eqs.~(S1)-(S10) therein]. The previous experiments of Refs.~\cite{Wei2022,Rosenberg2024} seemed to use $w_2(L,t)$. 

In the main text, we adopt $w_1(L,t)$ as the roughness $w(L,t)$. For a general initial state, it is better to use $w_2(L,t)$ as the definition of the roughness, but, as proved in the following, we have $w_1(L,t) = w_2(L,t)$ when the initial state is a pure Fock state. Hence, for such a pure Fock state, it is enough to consider $w_1(L,t)$, as used in Ref.~\cite{Fujimoto2020,Fujimoto2021,Fujimoto2022}. However, when the initial state is different from such a pure Fock state, the relation $w_1(L,t) = w_2(L,t)$ does not hold. In what follows, we explain this point in detail and then justify the use of $w_1(L,t)$ in our work. 

We shall prove $w_1(L,t) = w_2(L,t)$ when the initial state is a pure Fock state. We denote the initial pure Fock state by $\ket{a}$ ($\rho(0) = \ket{a} \bra{a}$). This is the eigenstate of $\hat{h}_1(0)$ and we denote the corresponding eigenvalue by $a$. Then, we can show 
\begin{align} 
 \bra{a} \hat{h}_2(t) \ket{a}    &= \bra{a} \hat{h}_1(t) \ket{a}  - \bra{a} \hat{h}_1(0) \ket{a}, \nonumber\\
 &= \bra{a} \hat{h}_1(t) \ket{a}  - a, \\
 \bra{a} \mathcal{T} \hat{h}_2(t)^2 \ket{a} &= \bra{a} \hat{h}_1(t)^2 \ket{a} - 2\bra{a} \hat{h}_1(t) \hat{h}_1(0) \ket{a}  + \bra{a} \hat{h}_1(0)^2 \ket{a} \nonumber\\
 &= \bra{a} \hat{h}_1(t)^2 \ket{a}  - 2 a \bra{a} \hat{h}_1(t) \ket{a} + a^2
\end{align}
Thus, we obtain
\begin{align} 
w_2(L,t)^2 
&= \bra{a} \mathcal{T} \hat{h}_2(t)^2 \ket{a} -  \bra{a} \hat{h}_2(t) \ket{a}^2 \nonumber \\
&= \bra{a} \hat{h}_1(t)^2 \ket{a} - \bra{a} \hat{h}_1(t) \ket{a}^2  \nonumber \\
&= w_1(L,t)^2. 
\end{align}
This completes the proof of $w_1(L,t) = w_2(L,t)$ for the initial pure Fock state. While we do not consider the ensemble average of the randomness of the kicked potential here, the straightforward calculation leads to the same relation even for such a case. 

When the initial state is a mixed Fock state, the term $\hat{h}_{1}(0)$ can contribute to the roughness dynamics, and the relation $w_1(L,t) = w_2(L,t)$ is broken down. Let us consider the following initial mixed Fock state:
\begin{align} 
\hat{\rho}(0) = \sum_a p_a \ket{a} \bra{a}. 
\label{rho}
\end{align}
Here, $p_a$ is the probability of finding the state $\ket{a}$ in the initial state, and the relation $\sum_a p_a = 1$ is satisfied. Under this setup, we can derive
\begin{align} 
{\rm Tr} \left( \hat{\rho}(0)  \hat{h}_2(t) \right)    
&= {\rm Tr} \left( \hat{\rho}(0) \hat{h}_{1}(t) \right)  - \sum_a p_a \bra{a} \hat{h}_{1}(0) \ket{a} \nonumber\\
&= {\rm Tr} \left( \hat{\rho}(0) \hat{h}_{1}(t) \right)  - \sum_a p_a a, \\
{\rm Tr} \left( \hat{\rho}(0)  \mathcal{T} \hat{h}_2(t)^2 \right) &= {\rm Tr} \left( \hat{\rho}(0) \hat{h}_2(t)^2 \right) \nonumber\\
&= {\rm Tr} \left( \hat{\rho}(0) \hat{h}_{1}(t)^2 \right)  - 2 \sum_a p_a a \bra{a} \hat{h}_{1}(t) \ket{a} +  \sum_a p_a a^2.
\end{align}
Hence, we can get
\begin{align} 
&w_2(L,t)^2  \\
&= {\rm Tr} \left( \hat{\rho}(0) \hat{h}_{1}(t)^2 \right)  - 2 \sum_a p_a a \bra{a} \hat{h}_{1}(t) \ket{a} + \sum_a p_a a^2   - \left(  {\rm Tr} \left( \hat{\rho}(0) \hat{h}_{1}(t) \right)  - \sum_a p_a a \right)^2 \nonumber\\
& =  w_1(L,t)^2  + w_1(L,0)^2 - 2 \sum_a p_a a \bra{a} \hat{h}_{1}(t) \ket{a} + 2 \left( \sum_a p_a a \right)  {\rm Tr} \left( \hat{\rho}(0) \hat{h}_{1}(t) \right).  \label{mix_h}
\end{align}
One readily finds that the last three terms on the right-hand side of Eq.~\eqref{mix_h} do not vanish for a $mixed$ initial state, while they vanish for an initial $pure$ Fock state ($p_a = 1$ for a single label $a$). This result means that the roughness dynamics can be affected by the existence of $\hat{h}_1(0)$ in  Eq.~\eqref{eq:h2} when the initial state is mixed. In the experimental situations, the prepared initial states are expected to be very close to pure Fock states, but there are small deviations from the pure state [see Eq.~\eqref{eq:initial1}]. Thus, we need to investigate the effect of a mixed state on roughness dynamics quantitatively.

\begin{figure}
\begin{center}
\includegraphics[width=0.9\linewidth]{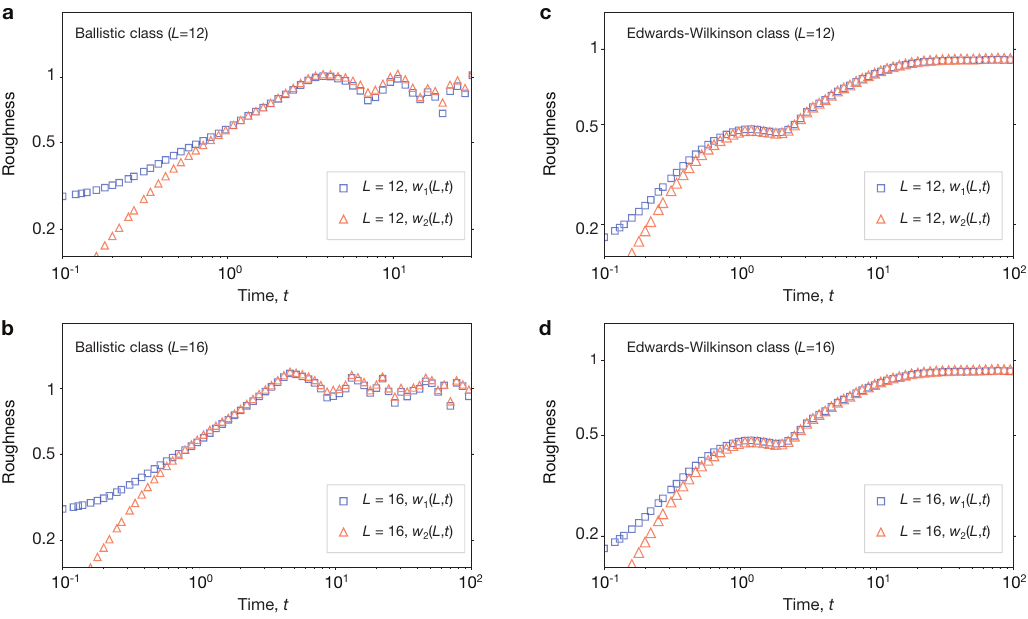}
\caption{
\textbf{Numerical results of surface roughness dynamics with different definitions.} 
\textbf{a,b,} Growth dynamics of $w_1(L,t)$ and $w_2(L,t)$ for the ballistic class and \textbf{c,d,} the Edwards-Wilkinson class.
The upper and lower panels display the results for $L=12$ and $ L=16$, respectively. The kicked potential used here is (V2) 4-site-bump potential. In all the cases, the differences between $w_1(L,t)$ and $w_2(L,t)$ appear in the short-time dynamics ($t < 1$), which is irrelevant for the universal nature of roughness dynamics. 
} 
\label{N_fig3} 
\end{center}
\end{figure}

We numerically calculate the difference between $w_1(L,t)$ and $w_2(L,t)$. The numerical setup is identical to that for the main text. Fig.~\ref{N_fig3} shows that the long-time dynamics for $w_1(L,t)$ and $w_2(L,t)$ are almost identical, while the short-time dynamics ($t<1$) exhibit some differences. Thus, the scaling property appearing in the long-time dynamics is robust against the effect of $\hat{h}_1(0)$ under the current experimental situations. This finding clearly justifies the use of $w_1(L,t)$ as the roughness in our experimental work.

\end{document}

%% file: preamble.tex
\usepackage{graphicx,
            caption,
            subcaption,
            amsmath,
            braket,
            amssymb,
            amsfonts,
            mathtools,
            ragged2e,
            xspace,
            nicefrac,
            lipsum,
            nameref,
            textcomp,
            urwchancal,
            float,
            upgreek,
            isotope,
            hyperref,
            cleveref,
            siunitx,
            placeins, 
            scrextend, 
            url}

\usepackage{gensymb}

\usepackage{scalerel}
\usepackage{tikz}
\usetikzlibrary{svg.path}

\DeclareCaptionLabelSeparator{dot}{. }
\makeatletter
\def\justified{
	\let\\\@normalcr
	\@rightskip\z@skip \rightskip\@rightskip
	\leftskip\z@skip
	\parindent 0em\relax
	\setlength{\parfillskip}{0pt plus 1fil}}
\DeclareCaptionJustification{justified}{\justified}
\hypersetup{colorlinks=true, linkcolor=blue,citecolor=blue,urlcolor=blue}
\def\unit #1 #2 {\SI{#1}{#2}\xspace}
\def\range #1 #2 #3 {\SIrange{#1}{#2}{#3}\xspace}
\DeclareSIUnit\gauss{G}

\newcommand{\myref}[2][]{Fig.~\hyperref[#2]{\ref*{#2}#1}}
\newcommand{\Myref}[2][]{Figure~\hyperref[#2]{\ref*{#2}#1}}
\newcommand{\Mytabref}[2][]{Table~\hyperref[#2]{\ref*{#2}#1}}

\usepackage{bm}
\usepackage{comment}
\usepackage{float} 
\usepackage[version=4]{mhchem} 

\definecolor{orcidlogocol}{HTML}{A6CE39}
\tikzset{
  orcidlogo/.pic={
    \fill[orcidlogocol] svg{M256,128c0,70.7-57.3,128-128,128C57.3,256,0,198.7,0,128C0,57.3,57.3,0,128,0C198.7,0,256,57.3,256,128z};
    \fill[white] svg{M86.3,186.2H70.9V79.1h15.4v48.4V186.2z}
                 svg{M108.9,79.1h41.6c39.6,0,57,28.3,57,53.6c0,27.5-21.5,53.6-56.8,53.6h-41.8V79.1z M124.3,172.4h24.5c34.9,0,42.9-26.5,42.9-39.7c0-21.5-13.7-39.7-43.7-39.7h-23.7V172.4z}
                 svg{M88.7,56.8c0,5.5-4.5,10.1-10.1,10.1c-5.6,0-10.1-4.6-10.1-10.1c0-5.6,4.5-10.1,10.1-10.1C84.2,46.7,88.7,51.3,88.7,56.8z};
  }
}

\newcommand\orcidicon[1]{\href{https://orcid.org/#1}{\mbox{\scalerel*{
\begin{tikzpicture}[yscale=-1,transform shape]
\pic{orcidlogo};
\end{tikzpicture}
}{|}}}}

%% file: authors_FVscaling.tex
\date{\today}

\newcommand{\kaist}[0]{\affiliation{Department of Physics, Korea Advanced Institute of Science and Technology, Daehak 291, Daejeon, 34141, Republic of Korea}}

\newcommand{\ist}[0]{\affiliation{Department of Physics, Institute of Science Tokyo, Tokyo, 152-8551, Japan}}

\newcommand{\pri}[0]{\affiliation{Nonequilibrium Quantum Statistical Mechanics RIKEN Hakubi Research Team, RIKEN Pioneering Research Institute (PRI), Saitama, 351-0198, Japan}}

\newcommand{\items}[0]{\affiliation{RIKEN Center for Interdisciplinary Theoretical and Mathematical Sciences (iTHEMS), RIKEN, Saitama, 351-0198, Japan}}

\newcommand{\nagoya}[0]{\affiliation{Department of Applied Physics, Nagoya University, Nagoya, 464-8603, Japan}}

\newcommand{\rccme}[0]{\affiliation{Research Center for Crystalline Materials Engineering, Nagoya University, Nagoya, 464-8603, Japan}}


\author{Kiryang\, Kwon\,}
\thanks{Current address: Massachusetts Institute of Technology, Cambridge, Massachusetts 02139, USA.}
\kaist

\author{Kazuya\, Fujimoto\,}
\ist

\author{Junhyeok\, Hur\,}
\kaist

\author{Byungjin\, Lee\,}
\kaist

\author{Samgyu\, Hwang\,}
\kaist

\author{Sumin\, Kim\,}
\kaist

\author{Ryusuke\, Hamazaki\,}
\pri
\items

\author{Yuki\, Kawaguchi\,}
\nagoya
\rccme

\author{Jae-yoon\,Choi\,}
\thanks{Email:  \mbox{\url{jaeyoon.choi@kaist.ac.kr}}}
\kaist